\documentclass[epj]{webofc}
\usepackage[utf8]{inputenc}
\usepackage[varg]{txfonts}   
\usepackage{booktabs}
\usepackage{xcolor}
\definecolor{darkred}{rgb}{0.4,0.0,0.0}
\definecolor{darkgreen}{rgb}{0.0,0.4,0.0}
\definecolor{darkblue}{rgb}{0.0,0.0,0.4}
\usepackage[bookmarks,linktocpage,colorlinks,
    linkcolor = darkred,
    urlcolor  = darkblue,
    citecolor = darkgreen]{hyperref}
\usepackage{subfigure}
\wocname{EPJ Web of Conferences}
\woctitle{Lattice2017}
\begin{document}
%
\selectlanguage{english}
\title{%
The topological properties of QCD at high temperature: \\
problems and perspectives
}
\author{%
\firstname{Claudio}  \lastname{Bonati}\inst{1}\fnsep\thanks{Present
address Dipartimento di Fisica dell'Universit\`a di Pisa and INFN - Sezione di
Pisa, Largo Pontecorvo 3, I-56127 Pisa, Italy,
\email{claudio.bonati@df.unipi.it}\ .}
}
\institute{%
Dipartimento di Fisica e Astronomia dell'Universit\`a di Firenze and INFN -
Sezione di Firenze, \\ Via Sansone 1, 50019, Sesto Fiorentino (FI), Italy.
}
\abstract{%
Lattice computations are the only first principle method capable of
quantitatively assessing the topological properties of QCD at high temperature,
however the numerical determination of the topological properties of QCD,
especially in the high temperature phase, is a notoriously difficult problem.
We will discuss the difficulties encountered in such a computation and some
strategies that have been proposed to avoid (or at least to alleviate) them.
}
\maketitle

\section{Introduction}\label{sec:intro}

Since the discovery of the instanton solution \cite{Belavin:1975fg} and the
introduction of the $\theta$-vacuum \cite{Callan:1976je, Jackiw:1976pf} there
has been constant interest in the study of the intrinsically non-perturbative
aspects of gauge theories related to the $\theta$-dependence. Such an interest
was motivated by phenomenological and theoretical reasons: from the theoretical
point of view $\theta$-dependence is a new knob that can be used to investigate
the properties of strongly interacting gauge theories, from the
phenomenological side it is connected with the anomalous $U_A(1)$ symmetry of
massless QCD and the $CP$ invariance of strong interactions.

The experimental fact that strong interactions are invariant under the discrete
$CP$ symmetry is indeed related to the unnaturally small value of the
(dimensionless, renormalization group invariant) $\theta$ parameter entering
the QCD lagrangian, with the best present estimate $|\theta|\lesssim 10^{-10}$
being obtained from measures of the neutron electric dipole moment
\cite{Afach:2015sja}. One of the most appealing way of explaining this small
value of $\theta$ is to assume it to be in fact exactly zero due to some new
symmetry. This approach was initiated by Peccei and Quinn \cite{Peccei:1977hh,
Peccei:1977ur} and, soon after its proposal, this mechanism was shown to imply
the existence of a new light pseudoscalar particle \cite{Weinberg:1977ma,
Wilczek:1977pj}, the axion, whose properties are strictly related to the 
behaviour of the effective potential for small $\theta$ values.

More or less during the same years the theoretical study of $\theta$
dependence in QCD was systematized and, in particular, it was realized that two
different behaviours are to be expected in the low and in the high temperature
regimes. At $T=0$ chiral perturbation theory can be used to compute the
effective potential as a function of the $\theta$ value
\cite{DiVecchia:1980yfw, diCortona:2015ldu}, obtaining at leading order
for the physical case of two light flavours the result
\begin{equation}\label{eq:chiPT}
V(\theta)=-m_{\pi}^2f_{\pi}^2\sqrt{1-\frac{4m_um_d}{(m_u+m_d)^2}\sin^2\frac{\theta}{2}}\ .
\end{equation}
In the very high temperature regime one can instead use perturbation theory
around instantons to show that the instanton density gets lower and lower as
the temperature increases, and this implies for the free energy density the
functional form \cite{Gross:1980br}
\begin{equation}\label{eq:f_DIGA}
f(\theta, T)_{DIGA}=f(0,T)+\chi_{DIGA}(T)(1-\cos\theta)\ ,
\end{equation}
where the subscript ``DIGA'' stands for Dilute Instanton Gas Approximation.
When $N_f$ light flavours are present, the coefficient $\chi_{DIGA}(T)$ scales
with the temperature as \cite{Gross:1980br}
\begin{equation}\label{eq:chi_DIGA}
\chi_{DIGA}(T)\propto m^{N_f}T^{4-\frac{11}{3}N_c-\frac{1}{3}N_f}\ .
\end{equation}

Finally, in \cite{Preskill:1982cy, Abbott:1982af, Dine:1982ah}, the axion field
was recognized to be a good cold dark matter candidate and it was realized that
this implies an overclosure bound: using the knowledge of the $\theta$
dependence of QCD at high temperature we can estimate the number of axions
produced in the ealry universe and thus the axion density in today universe.
From the fact that this density has to be smaller than the dark matter density
we obtain constraints on the axion interactions and, in particular, a lower
bound for its mass (modulo some cosmological hypotheses), see \cite{Moore} for
more details.  In order to have reliable bounds (or estimates, if we assume all
the dark matter to be composed of axions) we need to know with reasonable
accuracy the $\theta$ dependence of QCD at finite temperature and, in
particular, across the deconfinement/chiral restoration transition. The same
information is also needed if we are interested in studying the interplay
between topology and color confinement in gauge theories.

Since solid analytical results for the $\theta$ dependence are available
only for the cases of very low temperature (from chiral effective field
theory) and asymptotically high temperatures (from perturbation theory),
Lattice QCD is the only approach available to systematically investigate the
phenomena taking place for temperatures of the order of the critical
temperature $T_c$.  Some of the points we would like to clarify are:
\begin{itemize}
\item up to which temperature is it reasonable to use the predictions of chiral
perturbation theory?
\item what happens for $T$ close to $T_c$?
\item for which temperature we can start using DIGA results? 
\end{itemize}

In order to answer these questions it is convenient to use a general
parametrization of the $\theta$ dependence of the free energy density
$f$. Assuming $f(\theta, T)$ to be analytic in $\theta$ and using the
simple property $f(\theta, T)=f(-\theta, T)$ we can write the free energy
density in the general form \cite{Vicari:2008jw}
\begin{equation}\label{eq:gen_param}
f(\theta, T)=f(0,T)+\frac{1}{2}\chi(T)\theta^2\left(1+\sum_{n=1}^{\infty} b_{2n}(T)\theta^{2n}\right)\ .
\end{equation}
Since $\theta$ enters the Euclidean lagrangian only through
\begin{equation}
\mathcal{L}_{\theta}=-i\theta Q\ ,\quad Q=\int q(x)\mathrm{d}^4 x\ ,\quad q(x)=\frac{g^2}{64\pi^2}
\epsilon_{\mu\nu\rho\sigma}F^{a}_{\mu\nu}(x)F^{a}_{\rho\sigma}(x)\ ,
\end{equation}
the coefficients $\chi(T)$ and $b_{2n}(T)$ defined by Eq.~\eqref{eq:gen_param}
can be related to the cumulants of the topological charge $Q$ computed at
vanishing $\theta$; it is for example easy to show that the topological
susceptibility $\chi(T)$ and the coefficient $b_2(T)$ are given by the
expressions
\begin{equation}\label{eq:chi_b2}
\chi(T)=\frac{\langle Q^2\rangle_{\theta=0}}{\mathcal{V}},\quad 
b_2=-\frac{\langle Q^4\rangle_{\theta=0} -3\langle Q^2\rangle_{\theta=0}^2}{12\langle Q^2\rangle_{\theta=0}}\ ,
\end{equation}
where $\mathcal{V}$ is the four dimensional volume of the lattice in physical
units. At $T=0$ we have (for physical up and down quark masses) the next to
leading order $\chi$PT values $\chi(T=0)^{1/4}=75.5(5)~\mathrm{MeV}$ and
$b_2(T=0)=-0.029(2)$ (see \cite{diCortona:2015ldu}), while in the DIGA regime
it is immediate to get from Eq.~\eqref{eq:f_DIGA} the values $b_2=-1/12$ and
$b_4=1/360$.

\section{The problems}\label{sec:problems}

To investigate the $\theta$ dependence of QCD by means of lattice simulations
we have to evaluate, using Eq.~\eqref{eq:chi_b2} or equivalent forms, the
coefficients $\chi(T)$ and $b_{2n}(T)$ appearing in the general
parameterization of the free energy density in Eq.~\eqref{eq:gen_param}.  

As usual, the pure Yang-Mills theory is a convenient test bed to inquire in a
computationally simpler setup the level of accuracy that one can hope to reach
in QCD. The numerical study of the topological properties of Yang-Mills theory
at finite $T$ has quite a long story and has reached solid results on basically all
the open fronts. It is indeed known that 
\begin{itemize}
\item the topological susceptibility is practically independent of the
temperature up to deconfinement (see e.g.  \cite{Alles:1997qe} for an early
study),
\item close to $T_c$ the functional form of the $\theta$
dependence abruptly changes and for temperatures slightly above deconfinement
Eq.~\eqref{eq:f_DIGA} well describes data (see \cite{Bonati:2013tt} and later
improvements in \cite{Bonati:2015sqt,Bonati:2016tvi}),
\item the DIGA-like scaling of the topological susceptibility $\chi(T)\propto
T^{-7}$ starts for the $SU(3)$ case at surprisingly low temperatures
\cite{Berkowitz:2015aua, Kitano:2015fla, Borsanyi:2015cka} but DIGA predictions
underestimate $\chi(T)$ by about an order of magnitude in the temperature range
$T_c\lesssim T\lesssim 4T_c$.
\end{itemize}
It is important to note that these results are not easy to obtain even for the
pure glue case. The obstacles encountered are similar to the ones that
are found also in the QCD case, however in Yang-Mills theory, where more
efficient algorithms and higher statistics are available, it is possible to
proceed by brute force.  The two main problems encountered are the so called
freezing of the topological charge and the presence of particularly severe
finite size effects at high temperature.

\begin{figure}[b]
  \centering
  \includegraphics[width=0.45\textwidth, clip]{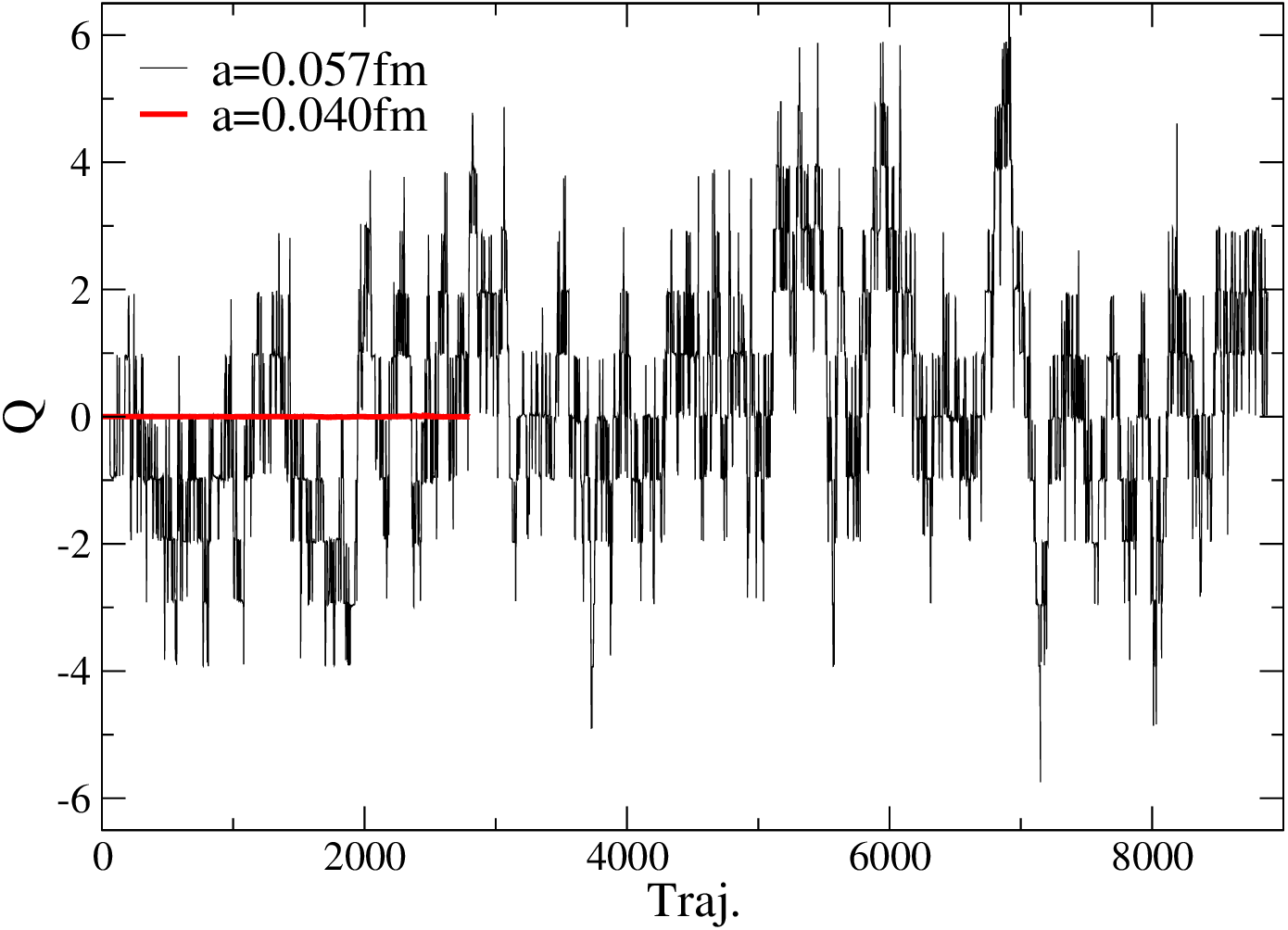}
  \hspace{0.5cm}
  \includegraphics[width=0.468\textwidth, clip]{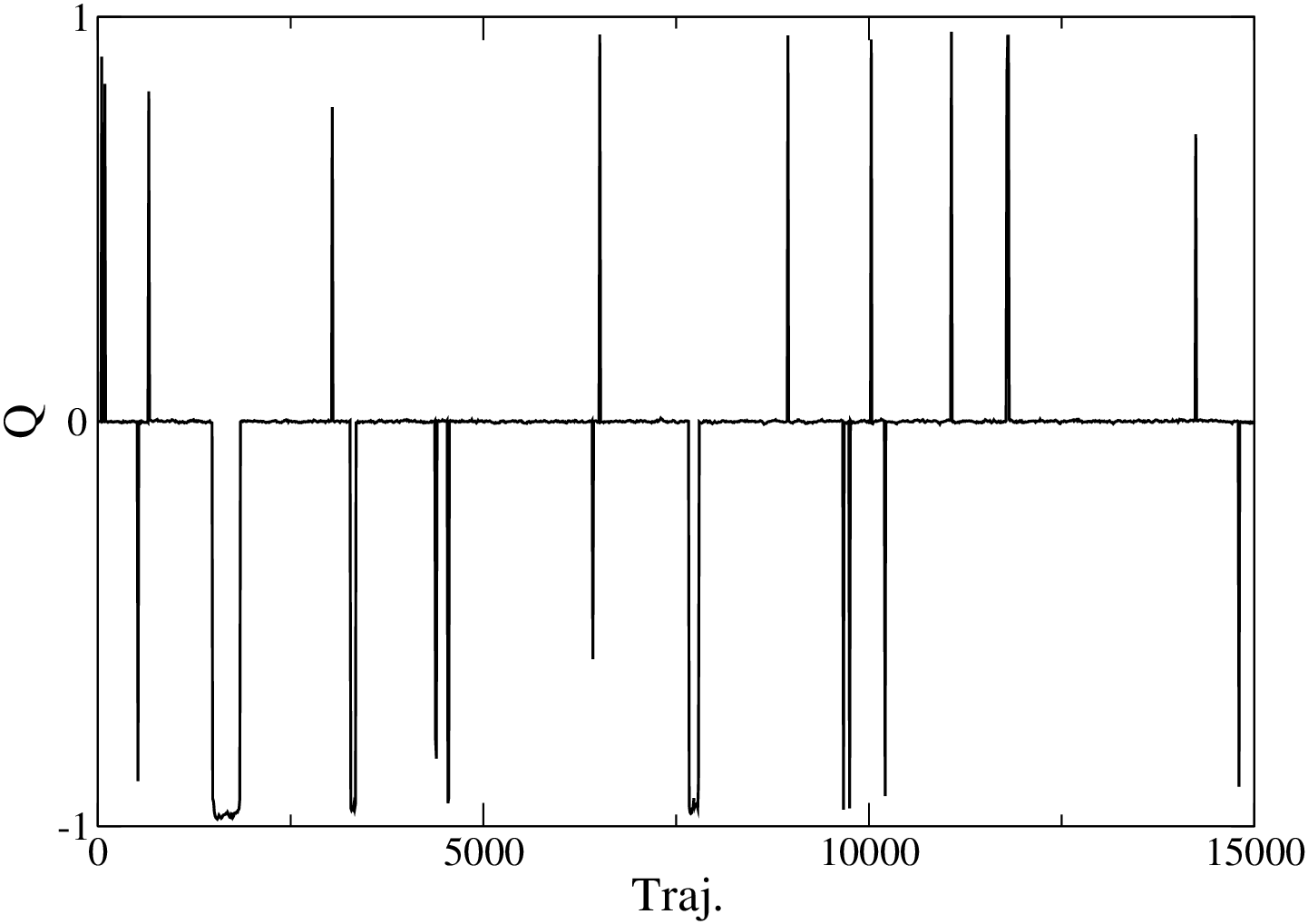}
  \caption{(left) Time histories of the topological charge $Q$ measured in a
  $N_f=2+1$ simulation with stout improved staggered fermions, tree-level
  Symanzik improved gauge action and physical quark masses on an hypercubic
  lattice of about $2~\mathrm{fm}$ extent. Passing from $a\simeq
  0.057~\mathrm{fm}$ to $a\simeq 0.040~\mathrm{fm}$ is sufficient for the
  topological charge to get stuck in the $0$ value for 3000 trajectories.  (right)
  History of $Q$ obtained by using the same discretization as in the left panel
  and $a\simeq 0.40~\mathrm{fm}$ but on a $16\times 64^3$ lattice, corresponding
  to $T\simeq 310~\mathrm{MeV}$. Configurations with zero topological charge are
  strongly favoured, with occasional spurts at $Q=\pm 1$ and some freezing
  visible at $Q=-1$.}
  \label{fig:problems}
\end{figure}

What goes under the name of ``freezing of the topological charge'' is the fact
that the autocorrelation time of the topological charge $Q$ quickly becomes
very large as the continuum limit is approached, until ergodicity is lost and
reliable simulations cannot be performed any more. Since $Q$ is an integer (or
almost integer, depending on the discretization adopted) number, in simulations
performed at (too) fine lattice spacing the topological charge stays constant
during all the simulation time, i.e. it is frozen. An example of this behaviour is
shown in Fig.~\ref{fig:problems}(left). 

The fact that the autocorrelation time of $Q$ gets larger and larger as the
continuum limit is approached is quite easy to understand: since $Q$ is a
topological invariant, to change its value it is necessary to pass through
configurations that are practically discontinuous and these configurations are
more and more suppressed as we get closer to the continuum limit.  Given this
simple argument we could guess the autocorrelation time of $Q$ to scale
approximately as the exponential of the correlation length in lattice units.
This scaling form is compatible with numerical results obtained in QCD
\cite{Alles:1996vn, Schaefer:2010hu} and in other systems with topologically
stable classical solutions \cite{DelDebbio:2002xa, DelDebbio:2004xh,
Campostrini:1992ar, Flynn:2015uma, Bonati:2017woi}, however it is notoriously
difficult to distinguish numerically an exponential behaviour from a power law
behaviour with very large exponent.  Despite the uncertainties related to the
precise functional form of the critical slowing down of the topological charge,
it is clear that it is much more severe than the slowing down observed for
``non topological'' observables, and this puts strong constraints on the
lattice spacing values that can be used in simulations.

In the high temperature phase the freezing problem is still present (although
it appears at different values of the lattice spacing for different
temperatures) however in this case there is also a new source of trouble: since
the topological susceptibility goes to zero at high temperature, by keeping the
physical volume fixed and increasing the temperature we are reducing the
typical amount of topological charge that is present in the lattice, which is
given by $\sqrt{\chi(T)\mathcal{V}}$ (see Eq.~\eqref{eq:chi_b2}). When this
number becomes much smaller than $1$ we are effectively sampling only the $Q=0$
topological sector of the theory and it is very difficult to reliably estimate
both $\chi$ and the $b_{2n}$ coefficients.  In Fig.~\ref{fig:problems}(right)
an example of this phenomenon is shown in a case in which it is still not
extreme, i.e. in which also the $Q=\pm 1$ sectors are occasionally explored.

From the practical point of view this problem is analogous to the freezing
problem, however these two problems have deeply different origins: freezing is
a purely algorithmic issue and it is at least conceivable that an algorithm
exists that completely solves it\footnote{Likely by means of non-local moves,
see \cite{Bonati:2017woi} for an example in a toy model.}, the other problem is
instead of physical origin and it is related to the fact that at high
temperature, on a finite volume, the statistical weight of the $Q=0$
configurations becomes overwhelming in the path-integral. As a consequence
freezing can happen for every value of the topological charge, while in the
high temperature phase only the probability of the $Q=0$ sector is enhanced.

The two problems mentioned so far are common to both QCD and pure gauge
theories, however when dynamical light fermions are present a new problem
emerges: the convergence to the continuum limit of the topological
susceptibility is extremely slow and very fine lattice spacings are needed to
extract the continuum limit, see \cite{Bonati:2015vqz, Petreczky:2016vrs,
Borsanyi:2016ksw}. The origin of this slow convergence rate to the continuum is
the strong connection between the $\theta$ dependence and the chiral properties
of the dynamical fermions, that can be seen as a consequence of the index
theorem and of the chiral anomaly. Since in most lattice discretizations the
chiral symmetry is at least partially broken, large lattice artefacts appear in
the $\theta$ dependence and, in particular, in the topological susceptibility. 

To summarize the computational difficulties: to obtain reliable results for the
$\theta$ dependence we have to perform simulations at very fine lattice
spacings, since otherwise huge lattice artefacts are present. On the other hand
we cannot use very fine lattice spacings since, beyond the usual computational
burden, we also have the topological freezing problem. All this remains true in
the high temperature phase, where a strong suppression of the $Q\neq 0$ sectors
is also present. Just to make things even worst the topological susceptibility
goes to zero as we lower the quark masses, and thus the high $T$ finite
size effect worsen when using dynamical light fermions.

\section{Possible solutions}\label{sec:solutions}

Of the three problems discussed in the previous section the one that attracted
most of the activity in the past was the freezing problem, likely because it
appears not only in QCD but in all field theories (or even quantum mechanical
models) whose configuration space can be decomposed, in the continuum, into
homotopy classes.  This property makes the freezing problem amenable of being
studied using models of increasing computational complexity before attacking
the QCD case; typical examples of systems used in test studies are: quantum
mechanics on a circumference, two dimensional $CP^N$ models and Yang-Mills
theories.  In the years several methods have been developed to cope with the
freezing problem and two main approaches can be found in the literature: the
goal of the first group of algorithms is to reduce the autocorrelation time of
the topological charge, while methods of the second group aim at extracting as
much information as possible from frozen simulations.

One of the most widely used method of the first group is the use of open
boundary conditions \cite{Luscher:2011kk}. The idea of the method is quite
intuitive and it is related to the fact that with periodic boundary conditions
(i.e. effectively no boundaries) the only possible way of changing the topology
is to pass through discontinuous configurations having very small probability
in the path-integral; if instead we use open boundary conditions there can be
an inflow/outflow of topology from the boundary and the autocorrelation time of
$Q$ is expected to scale as $\propto a^{-2}$ \cite{McGlynn:2014bxa}.
Simulations adopting open boundary conditions has been by now used mainly to
study the $T=0$ case, in which it is natural to ``open'' the torus in the
temporal direction, however it is in principle possible to use the same idea
also at finite temperature, using open boundary conditions in a spatial direction
\cite{Florio}.  An obvious drawback of this method is however the fact that we
loose translation invariance\footnote{In order to reduce the breaking of the
translation invariance the use of a non-orientable manifold was proposed in
\cite{Mages:2015scv}.} and we have to study only the part of the lattice
that is far enough from the boundaries not to be significantly affected by
them.  In particular, expressions like Eq.~\eqref{eq:chi_b2} cannot
be used anymore, and the topological susceptibility has to be computed as the
integral of the two point function of the topological charge density $q(x)$,
with the two point function being estimated in the bulk of the lattice. The $b_{2n}$
coefficients can be in principle computed in an analogous way, using the
$2n-$point functions of $q(x)$, however this becomes increasingly difficult
for larger $n$ values.

Another method to improve the decorrelation of the topological charge is the
use of extended space approaches, like e.g. tempering methods. These have been
introduced in the study of spin glass systems and the idea is to mix together
in a stochastically exact combination simulations performed at different values
of the parameters, in such a way that the slowly decorrelating runs are speed
up by the quickly decorrelating ones. Two different practical implementations
of this general philosophy are simulated tempering \cite{Marinari:1992qd} and
parallel tempering \cite{HukushimaNemoto}.  In the spin glass case the value of
the autocorrelation time increases at low temperature and the parameter that is
used in the tempering procedure is thus temperature; for the case of the
freezing problem a natural choice is the lattice spacing. This approach has
been tested in toy models with promising results (see \cite{Vicari:1992jy} for
$CP^N$ models and \cite{Bonati:2017woi} for a quantum mechanical model), but a
systematical investigation for the case of lattice gauge theories has not yet
been performed. A different application of the tempering approach has been
advocated in \cite{Hasenbusch:2017unr}, where it has been shown (for the case
of $CP^N$ models) that parallel tempering can be efficiently used to connect
simulations performed using, instead of different parameter values, different
boundary conditions, in order to reduce the finite size effects of the open
boundary conditions while retaining some of their advantages.

A third class of approaches that can be used to reduce the autocorrelation time
of $Q$ in simulations is to sample, instead of the original distribution, a
different one that is computationally less problematic and finally reweight the
results; this is the idea of the multicanonical algorithm, introduced in
\cite{Berg:1992qua} for the study of strong first order transitions.
Multicanonical simulations are usually carried out in such a way that
configurations are sampled with a probability proportional to the inverse of
the density of states (of the original system), in order to obtain a flat
histogram in the modified system. As it always happens, there are several ways
of implementing this general idea. The original proposal \cite{Berg:1992qua}
was to use a rough estimate of the density of states, since the algorithm is
stochastically exact anyway, however it was later realized that some
improvements are possible on this basic approach. A possibility that can be
thought as a dynamical implementation of the multicanonical idea is
metadynamics \cite{LaioParr}, in which an history dependent biasing potential
is added to the system and updated in such a way that, as the simulation goes
on, the biasing potential converges to the logarithm of the density of states.
This idea has been successfully applied to the case of $CP^N$ models in
\cite{Laio:2015era} and we will discuss some more details of metadinamics in
the next section, where the application to the QCD case will be analyzed.  A
more radical approach is to precisely estimate the density of states
\cite{WangLandau, Langfeld:2012ah} and then compute observables without the
need of a Monte Carlo.  First results for the topological susceptibility
obtained in this way have been presented in \cite{Rago}.

As anticipated before, there are also algorithmic developments that go in the
direction of obtaining as much information as possible from simulations in
which the topology is frozen. An idea that is around since some time
\cite{Shuryak:1994rr} but that recently draw new attention is the use of
sub-volume measurements: even if the global topological charge is fixed we can
try to extract information from its local fluctuations.  In
\cite{Brower:2014bqa} a Maximum-Likelihood method was advocated to estimate the
parameters of the $\theta$ dependence, while the approach proposed in
\cite{Bietenholz:2015rsa} is more direct but, in its original formulation,
it is limited to the case in which the $b_{2n}$ coefficients vanish. It is
nevertheless reasonable to expect that, by sacrificing part of its simplicity, a
generalization of this algorithm to the case $b_{2n}\neq 0$ can be found.
Another interesting possibility, that was originally introduced for the case of
overlap fermions simulations, is to extract the topological susceptibility from
finite size effects of simulations performed at fixed $Q$ \cite{Brower:2003yx,
Aoki:2007ka}. An algorithm that uses also simulations at fixed topology is the
one introduced in \cite{Frison:2016vuc, Borsanyi:2016ksw} (see also
\cite{Kovacs}): if for some values of the parameters (e.g. at coarse lattice
spacing) it is possible to estimate the relative weights of the different
topological sectors, then it is possible to propagate this information to other
parameter values using an analogous of the thermodynamic integration for
$Z_Q/Z_0$, where $Z_Q$ is the partition function at fixed $Q$ value. From this
information is then possible to evaluate both the topological susceptibility
and the $b_{2n}$ coefficients, the only possible difficulty being the fact that
a large number of simulations could be needed if $\langle Q^2\rangle\gg 1$.

As previously explained in Sec.~\ref{sec:problems}, in simulations performed in
QCD with dynamical light flavours the freezing of the topological charge is a
particularly severe problem because of the fact that the topological
susceptibility converges to its continuum limit very slowly. An approach that
is complementary to the development of algorithms to cope with the freezing
problem is thus the study of techniques to speed up the approach to the
continuum limit for topological observables.  Once realized that the slow
convergence to the continuum can be related to the lattice breaking of the
chiral symmetry, it is natural to investigate the possibility of correcting a
posteriori this breaking in the numerical results to improve the convergence. 

In the $T=0$ case we can rely on chiral perturbation theory at finite lattice
spacing in order to have a better understanding of the observed numerical
behaviour. Staggered fermions have been up to now the most widely used
discretization in studies of the $\theta$ dependence\footnote{For an exception
to this statement see e.g. \cite{Trunin:2015yda, Burger:2017xkz}, where a
twisted mass Wilson fermion discretization was adopted.} and a staggered chiral
perturbation analysis of the topological susceptibility has been carried out in
\cite{Billeter:2004wx}. The outcome of this analysis is that, at non-vanishing
lattice spacing, the dominant violation of the expected chiral behaviour of the
topological susceptibility can be described by using the taste-singlet pion
mass $m_{\pi,\mathrm{ts}}$ instead of the physical pion mass $m_{\pi}$.  The
idea of correcting the numerical results by a factor related to the taste
splitting was introduced in \cite{Bonati:2015vqz} and more thoroughly
investigated in \cite{Borsanyi:2016ksw} and it was shown to be very effective
in improving the convergence to the continuum limit of the topological
susceptibility at zero temperature.

The finite temperature case is more problematic, since there are no solid
theoretical results available for generic temperatures. In
\cite{Bonati:2015vqz} it was proposed to use the dimensionless ratio $\chi(T,
a)/\chi(T=0, a)$, with the idea that a large part of the lattice artefacts
would cancel in this ratio. This appeared to be the case for the lattice
spacings explored in \cite{Bonati:2015vqz}, however it was later shown in
\cite{Petreczky:2016vrs, Borsanyi:2016ksw} that the problem of the slow
convergence to the continuum is not solved using this recipe: when going to
still smaller lattice spacings the ratio $\chi(T, a)/\chi(T=0, a)$ starts again
to show a strong dependence on the value of $a$. In the Dilute Instanton Gas
Approximation each instanton (or anti-instanton) is associated to (at least) a
zero mode of the Dirac operator and, since instantons are weakly interacting by
hypothesis, the mass dependence of the partition function in the chiral limit
is dominated by these isolated zero modes. The approach proposed in
\cite{Borsanyi:2016ksw} (see also \cite{Kovacs}) consists in reweighting gauge
configurations in order to fix this behaviour. Let's take the example of a
$Q=1$ configuration: there should be at least a zero mode of the Dirac
operator\footnote{When using staggered fermions a larger number of zeroes has
to be present, however here we only want to present the general idea, for
details see \cite{Borsanyi:2016ksw}.} however, due to the explicit lattice
breaking of the chiral symmetry, only a quasi zero mode with eigenvalue
$\Delta$ will be present, in which case a weight
$\left(\frac{m}{m+\Delta}\right)^{N_f}$ is assigned to the configuration, where
$m$ is the light quark mass. Using this approach the value of the topological
susceptibility that is obtained in the high temperature phase is reduced by
more than an order of magnitude and the dependence of the result on the lattice
spacing is very mild. 

Let us close this section with some comments on the finite volume problem at
high temperature. Since for large enough temperature the system is practically
always in the charge zero sector, all the methods that have been developed to
extract information from frozen simulations are in principle applicable also to
this case. In particular the ``thermodynamic integration'' method of
\cite{Frison:2016vuc, Borsanyi:2016ksw} has been originally introduced in this
context and not as a solution to the freezing problem.  Also the approaches
that have been broadly classified before as ``multicanonical like'' (i.e.
multicanonical, metadynamics and density of states) can be profitably used at
high temperature, see the next section and \cite{Bonati:2017woi}.

\section{Some results from metadynamics}

In this section we will present the outcome of some tests carried out in
collaboration with M.~D'Elia, G.~Martinelli, F.~Negro and F.~Sanfilippo to
investigate the possibility of using metadynamics in the high temperature phase
of QCD. 

Metadynamics has been introduced to efficiently sample the configuration space
of models with several free energy minima for which we have been able to
identify (at least approximately) the slow degrees of freedom.  From this point
of view metadynamics is less versatile than e. g. tempering methods, however,
when it can be applied, it is expected to be more efficient, since it
explicitly uses more information on the system. The general philosophy of
metadynamics is the following \cite{LaioParr} (see \cite{LaioGervasio} for a
review)
\begin{enumerate}
\item identify the ``slow variable'' $\phi$, typically called ``collective
variable'' in the original literature (note that several slow variables can be
used together)
\item introduce in the action an external history-dependent potential $V(\phi,
\tau)$ (called metapotential) for the slow variable, where $\tau$ is the Monte
Carlo time
\item  during the update modify $V(\phi, \tau)$ in such a way that the
distribution of $\phi$ is (on average) constant in the interval of interest
$[\phi_{min},\phi_{max}]$
\item at the end of the simulation reweight data using $\langle V(\phi,
\tau)\rangle$, that is an estimator of the free energy at fixed $\phi$.
\end{enumerate}

The point 3) is typically implemented by increasing the value of the metapotential
corresponding to the value of $\phi$ that is currently in use, in order
to escape from free energy minima and explore uniformly the interval
$[\phi_{min},\phi_{max}]$.  As noted in the previous section this approach can
be intuitively thought as a dynamical multicanonical update, since we start
from a very rough estimate of the metapotential (like e.g. a flat distribution)
that is then self-consistently improved during the simulation. Metadynamics,
being a random walk biased toward exploring new values, is expected to converge
faster than the multicanonical update, that corresponds to a simple random
walk; on the other hand  the algorithm does not use anymore a stationary
Markov chain and, as a consequence, it is non trivial to prove its stochastic
exactness \cite{CrespoEtAl}.

\begin{figure}[b]
  \centering
  \includegraphics[width=0.4\textwidth, clip]{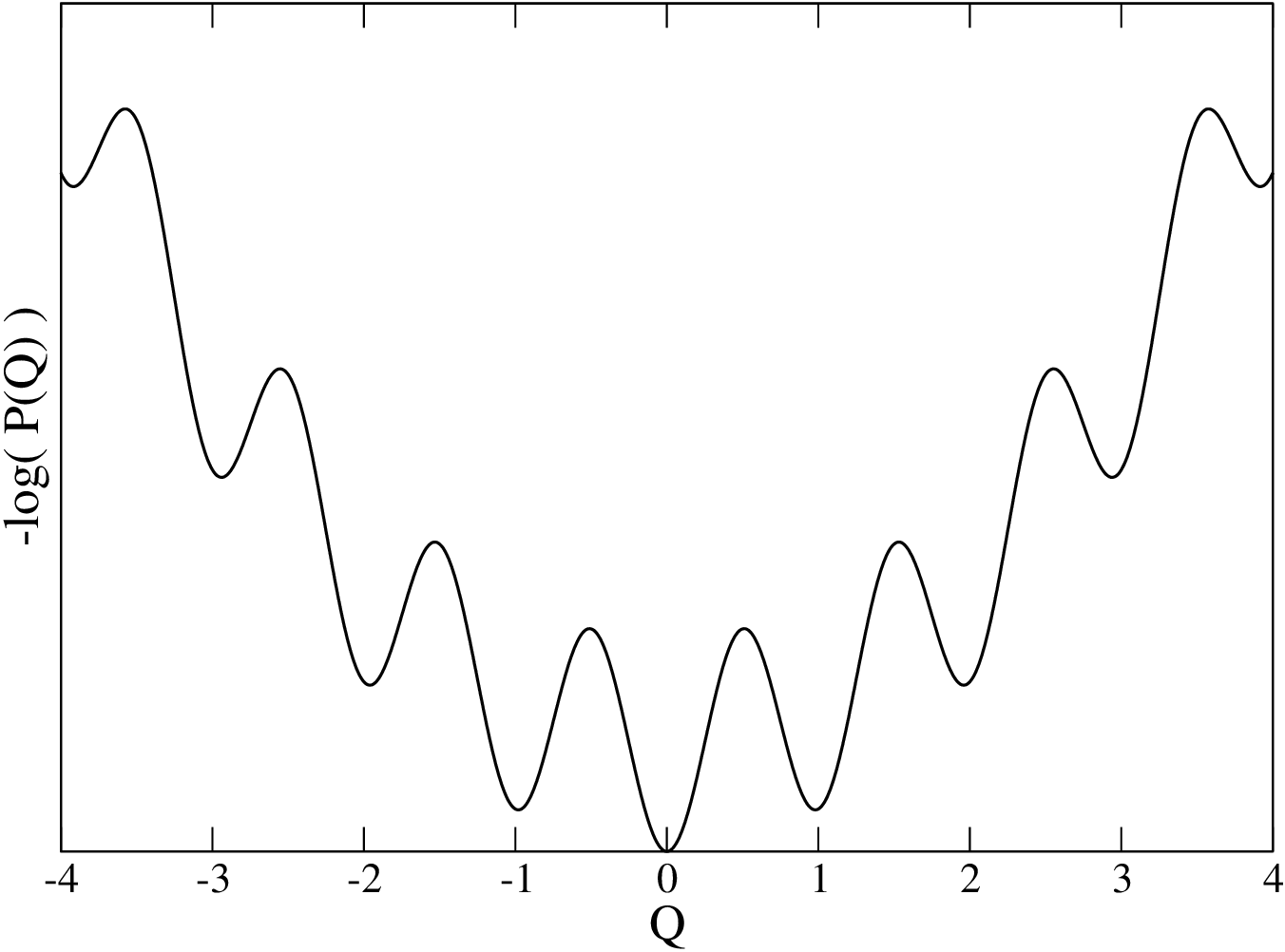}
  \hspace{0.5cm}
  \includegraphics[width=0.4\textwidth, clip]{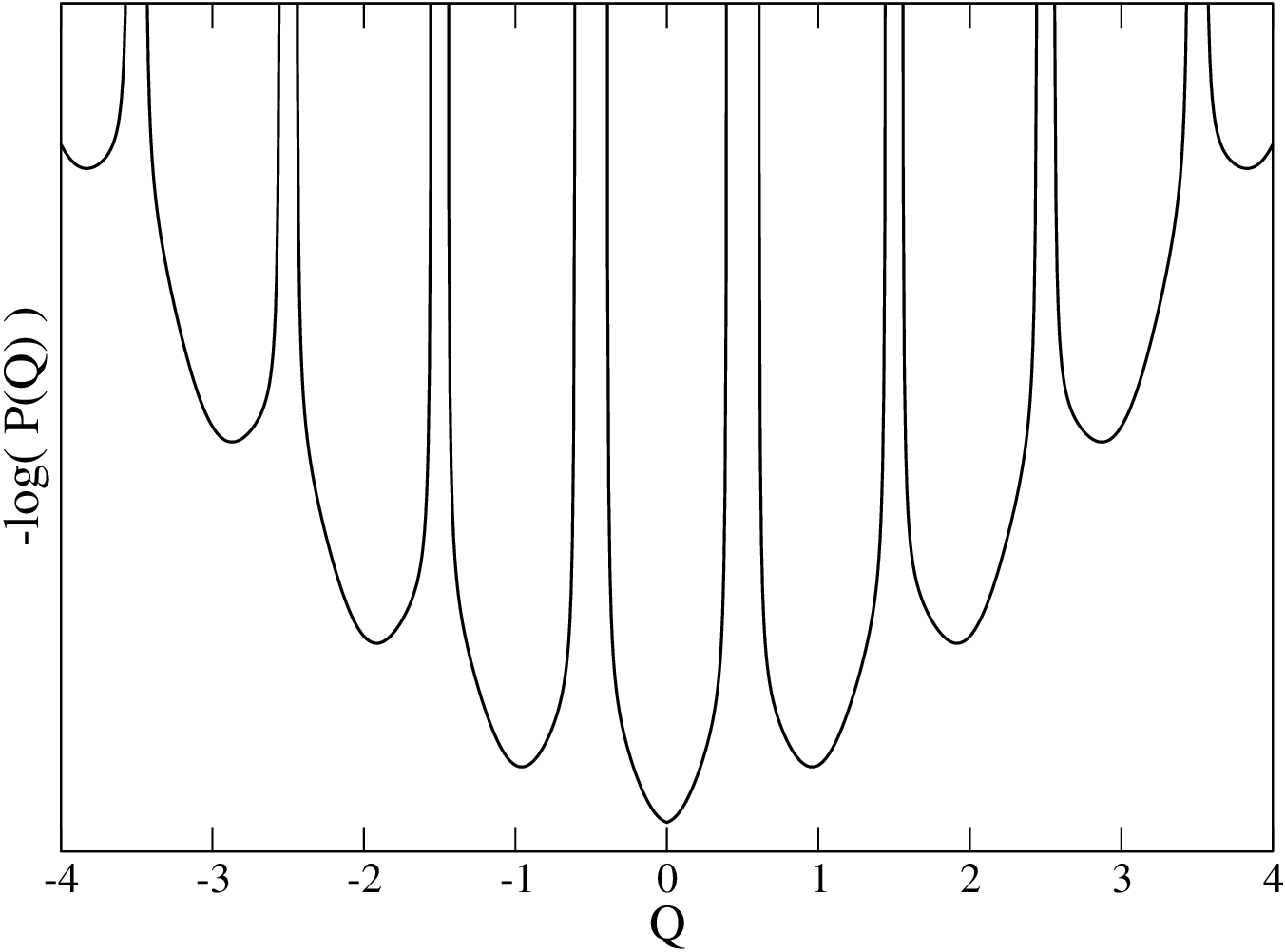}
  \caption{(left) A typical free energy at fixed $Q$ landscape ($P(Q)$ is the
  density of states) for which metadynamics is expected to work well.  (right)
  The free energy landscape that has to be expected for the topological charge
  $Q$.}
  \label{fig:naivelessnaive}
\end{figure}

In order to apply metadynamics we have first of all to identify the appropriate
collective variable to be used. A natural choice would be the topological
charge $Q$, however this would not work. If we use a discretization that gives
an exactly integer value for $Q$, like the geometric or fermionic definitions, the
metapotential induces no force in the evolution unless there is a topology
change in the trajectory; as a consequence the ``bias'' would not be effective
in the dynamics. Also the use of a definition of the topological charge that is
not integer by construction, like the one obtained by using the clover
discretization after having applied cooling or gradient flow, would not be
appropriate. Indeed this topological charge is no more exactly integer, but it is
still so strongly peaked at integer values that it would not sufficiently bias
the dynamics. A pictorial representation of this fact is reported in
Fig.~\ref{fig:naivelessnaive}. In Fig.~\ref{fig:naivelessnaive}(left) a free
energy profile is sketched for which the application of metadynamics is
expected to be useful: updating the metapotential we will gradually fill the local minima
and the whole configuration space will then be sampled. If we use
the topological charge $Q$ as collective variable the situation is however more
similar to Fig.~\ref{fig:naivelessnaive}(right): even using metadynamics it
would be difficult to fill the minima and escape the freezing.

From the previous analysis it emerges that a good collective variable should be
something that has good correlation with the topological charge $Q$ but that
can assume also far from integer values. Our choice was to use what could be
called an under-smoothed topological charge, i.e. the result obtained by
computing the clover discretization of the topological charge after just a few
smearing steps. We will denote this variable by $Q'$ and, more precisely, it is
defined as the value obtained by using the clover discretization of the
topological charge after having applied 20 stout smearing steps
\cite{Morningstar:2003gk} to the configuration, with smearing parameter
$\alpha=0.1$. While a priori other smoothing procedures can be adopted as well
in the definition of $Q'$, the use of a differentiable smearing is fundamental
for the implementation of metadynamics in a standard HMC code. 

A test run was performed by applying metadynamics to the collective variable
$Q'$ in a simulation with $N_f=2+1$ flavours of stout smeared staggered quarks
with physical quark masses and Symanzik tree level improved gauge action.  The
physical parameters adopted were the same as for the time history shown in
Fig.~\ref{fig:problems}(right), corresponding to $a\simeq 0.40\,\mathrm{fm}$
and $T\simeq 310\,\mathrm{MeV}$, and metadynamics was used to improve the
sampling of the interval $[Q'_{min},Q'_{max}]=[-3:3]$, i.e. the metapotential
takes the constant values $V(Q'=3,\tau)$ for $Q'>3$ and $V(Q'=-3,\tau)$ for
$Q'<-3$.

\begin{figure}[t]
  \centering
  \includegraphics[width=0.8\textwidth, clip]{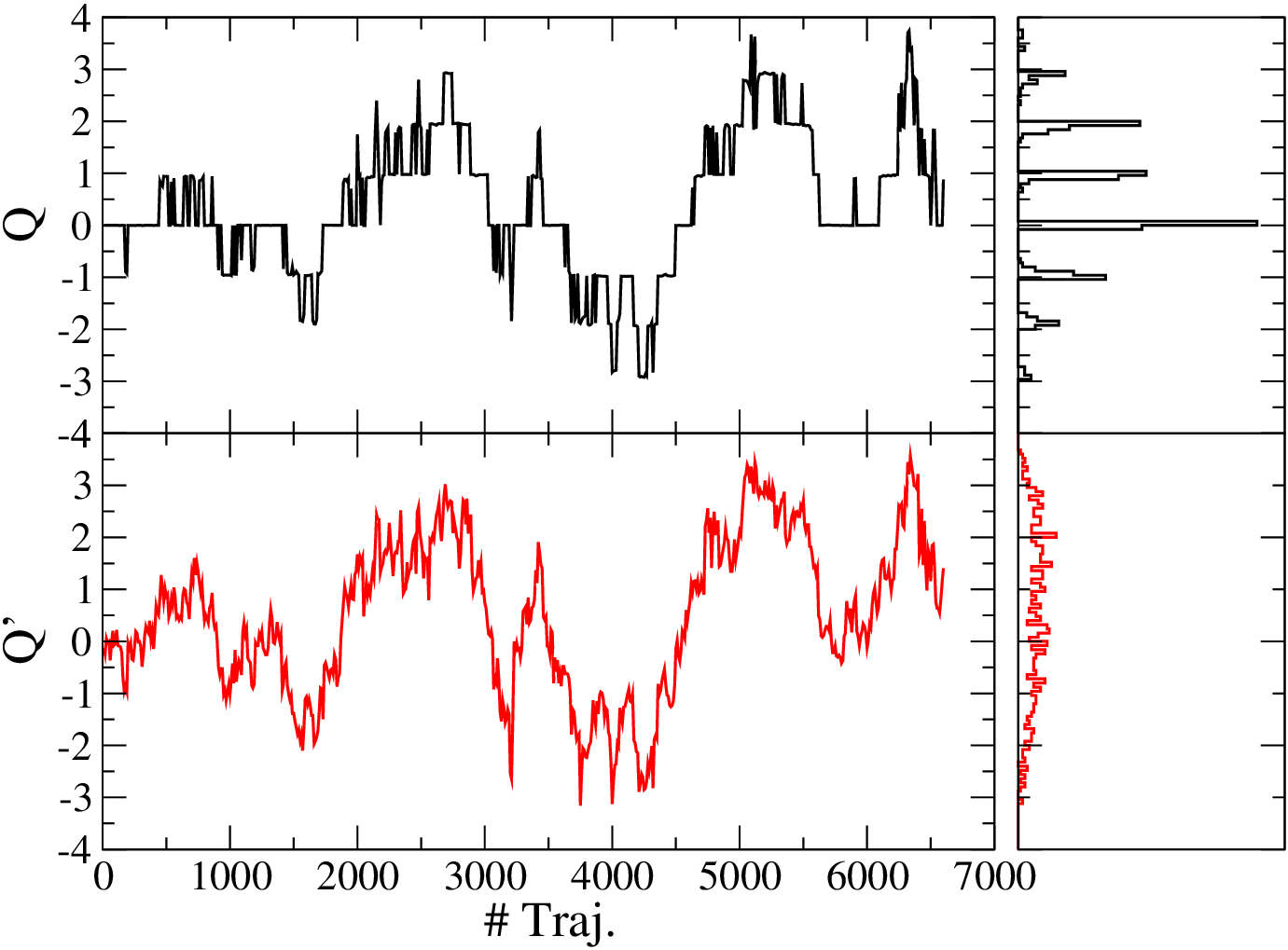}
  \caption{Time histories and histograms of the topological charge $Q$ and of the
under-smoothed topological charge $Q'$ obtained by using metadynamics. The same
physical parameters as in  Fig.~\ref{fig:problems}(right) were used.} 
  \label{fig:meta}
\end{figure}

In Fig.~\ref{fig:meta} the time histories of $Q$ and $Q'$ in the metadynamics
run are shown, together with their histograms, and several things can be noted:
first of all it is clear that there is a good correlation between $Q'$ and the
topological charge $Q$ measured after 70 cooling steps\footnote{We verified
that 70 cooling steps are sufficient to reach the plateau region in which the
value of the topological charge is almost independent of the number of cooling
steps used.}, moreover the histograms show that $Q$ is strongly peaked on
nearly integer values, while this is not the case for $Q'$, whose histogram is
nearly flat due to metadynamics. These were the requirements for the biasing
potential to be effective and indeed we can see that, notwithstanding the
temperature of about $310\,\mathrm{MeV}$ and the small lattice spacing
($a\simeq 0.04\,\mathrm{fm}$), all the topological sectors in the range
$[-3,3]$ are explored and no signal of freezing is present. The overhead
introduced by metadynamics is not negligible, since the time required to evolve
a trajectory when metadynamics is used is about three times the one required in
a standard HMC simulation; given the obtained results this seems however totally
worth it.

\section{Conclusions}

The study of the topological properties of non abelian gauge theories, and QCD
in particular, using lattice simulations has by now a long tradition, dating
back to early eighties \cite{DiVecchia:1981aev}.  At low temperature chiral
perturbation theory can be used, for very high $T$ perturbation theory and
instanton calculus are available, but the lattice approach is still the only
tool that can be used to investigate, from first principles and in a systematically
improvable way, the $\theta$ dependence of QCD for generic values of the
temperature. 

The fact that lattice simulations can (or have to) be used does not mean that
the numerical determination by lattice methods of the $\theta$ dependence is an
easy task, and in fact it is a surprisingly hard one. Mainly for this reason
most studies focused in the past on the pure glue case, which is also
problematic but for which a brute force approach can be pursued with success.
The recent phenomenological interest in axion properties triggered new lattice
studies of $\theta$ dependence in QCD with physical quark masses, that came
together with the developments of new algorithms.

The main problems that are encountered when trying to estimate the $\theta$
dependence of QCD with physical quark masses are the exceptionally large
autocorrelation time of the topological charge at small lattice spacing (the so
called ``freezing problem''), the fact that at high temperature the weight in
the path-integral of the $Q=0$ sector becomes overwhelming, the exceedingly
slow convergence to the continuum limit of the topological susceptibility (and
maybe also of higher cumulants of the topological charge).

The freezing problem was the one that attracted most of the algorithmic
developments, likely because it is not only typical of QCD but it appears in
all theories with topologically stable classical solutions. Nevertheless,
in the last couple of years, several new algorithms have been developed to cope
with all the previously mentioned problems and we are now in position to
crosscheck the different approaches that have been developed, in order to
systematically inquire the possible presence of systematics and provide solid
results for the QCD case. 


\section*{\small Acknowledgements}

It is a great pleasure to thank Massimo D'Elia, Guido Martinelli, Michele
Mesiti, Francesco Negro, Francesco Sanfilippo and Ettore Vicari for countless
discussions on the topics covered in this proceeding.  We acknowledge PRACE for
awarding us access to the Marconi supercomputer at CINECA (Italy) under project
Pra13\_3331.


\end{document}